\documentstyle[12pt,psfig]{article}
\makeatletter
\def\@cite#1#2{{[{#1}]\if@tempswa\typeout
{IJCGA warning: optional citation argument
ignored: `#2'} \fi}}

\newcount\@tempcntc
\def\@citex[#1]#2{\if@filesw\immediate\write\@auxout{\string\citation{#2}}\fi
  \@tempcnta\z@\@tempcntb\m@ne\def\@citea{}\@cite{\@for\@citeb:=#2\do
    {\@ifundefined
       {b@\@citeb}{\@citeo\@tempcntb\m@ne\@citea\def\@citea{,}{\bf ?}\@warning
       {Citation `\@citeb' on page \thepage \space undefined}}%
    {\setbox\z@\hbox{\global\@tempcntc0\csname b@\@citeb\endcsname\relax}%
     \ifnum\@tempcntc=\z@ \@citeo\@tempcntb\m@ne
       \@citea\def\@citea{,}\hbox{\csname b@\@citeb\endcsname}%
     \else
      \advance\@tempcntb\@ne
      \ifnum\@tempcntb=\@tempcntc
      \else\advance\@tempcntb\m@ne\@citeo
      \@tempcnta\@tempcntc\@tempcntb\@tempcntc\fi\fi}}\@citeo}{#1}}
\def\@citeo{\ifnum\@tempcnta>\@tempcntb\else\@citea\def\@citea{,}%
  \ifnum\@tempcnta=\@tempcntb\the\@tempcnta\else
   {\advance\@tempcnta\@ne\ifnum\@tempcnta=\@tempcntb \else \def\@citea{--}\fi
    \advance\@tempcnta\m@ne\the\@tempcnta\@citea\the\@tempcntb}\fi\fi}
\makeatother
\newenvironment{Eqnarray}%
     {\arraycolsep 0.14em\begin{eqnarray}}{\end{eqnarray}}

\def\simlt{\stackrel{<}{{}_\sim}}
\def\simgt{\stackrel{>}{{}_\sim}}
\def\be{\begin{equation}}
\def\ee{\end{equation}}
\def\bear{\be\begin{array}}
\def\eear{\end{array}\ee}
\def\bea{\begin{Eqnarray}}
\def\eea{\end{Eqnarray}}

\def\lsim{\mathrel{\raise.3ex\hbox{$<$\kern-.75em\lower1ex\hbox{$\sim$}}}}
\def\gsim{\mathrel{\raise.3ex\hbox{$>$\kern-.75em\lower1ex\hbox{$\sim$}}}}
\def\ifmath#1{\relax\ifmmode #1\else $#1$\fi}
\def\ls#1{\ifmath{_{\lower1.5pt\hbox{$\scriptstyle #1$}}}}

\def\beq{\begin{equation}}
\def\eeq{\end{equation}}
\def\beqa{\begin{Eqnarray}}
\def\eeqa{\end{Eqnarray}}

\def\snu{\tilde{\nu}}
\def\gappeq{\mathrel{\rlap {\raise.5ex\hbox{$>$}}
{\lower.5ex\hbox{$\sim$}}}}
\def\lappeq{\mathrel{\rlap{\raise.5ex\hbox{$<$}}
{\lower.5ex\hbox{$\sim$}}}}

\begin{document}
\def\IJMPA #1 #2 #3 {{\sl Int.~J.~Mod.~Phys.}~{\bf A#1}\ (19#2) #3$\,$}
\def\MPLA #1 #2 #3 {{\sl Mod.~Phys.~Lett.}~{\bf A#1}\ (19#2) #3$\,$}
\def\NPB #1 #2 #3 {{\sl Nucl.~Phys.}~{\bf B#1}\ (19#2) #3$\,$}
\def\PLB #1 #2 #3 {{\sl Phys.~Lett.}~{\bf B#1}\ (19#2) #3$\,$}
\def\PR #1 #2 #3 {{\sl Phys.~Rep.}~{\bf#1}\ (19#2) #3$\,$}
\def\JHEP #1 #2 #3 {{\sl JHEP}~{\bf #1}~(19#2)~#3$\,$}
\def\PRD #1 #2 #3 {{\sl Phys.~Rev.}~{\bf D#1}\ (19#2) #3$\,$}
\def\PTP #1 #2 #3 {{\sl Prog.~Theor.~Phys.}~{\bf #1}\ (19#2) #3$\,$}
\def\PRL #1 #2 #3 {{\sl Phys.~Rev.~Lett.}~{\bf#1}\ (19#2) #3$\,$}
\def\RMP #1 #2 #3 {{\sl Rev.~Mod.~Phys.}~{\bf#1}\ (19#2) #3$\,$}
\def\ZPC #1 #2 #3 {{\sl Z.~Phys.}~{\bf C#1}\ (19#2) #3$\,$}
\def\PPNP#1 #2 #3 {{\sl Prog. Part. Nucl. Phys. }{\bf #1} (#2) #3$\,$}

\catcode`@=11
\newtoks\@stequation
\def\subequations{\refstepcounter{equation}%
\edef\@savedequation{\the\c@equation}%
  \@stequation=\expandafter{\theequation}
  \edef\@savedtheequation{\the\@stequation}
  \edef\oldtheequation{\theequation}%
  \setcounter{equation}{0}%
  \def\theequation{\oldtheequation\alph{equation}}}
\def\endsubequations{\setcounter{equation}{\@savedequation}%
  \@stequation=\expandafter{\@savedtheequation}%
  \edef\theequation{\the\@stequation}\global\@ignoretrue

\noindent}
\catcode`@=12

\vspace*{-1in}
\renewcommand{\thefootnote}{\fnsymbol{footnote}}
\begin{flushright}
OUTP-01-22P \\
\end{flushright}
\vskip 5pt
\begin{center}
{\Large {\bf Determining See-Saw Parameters \\
from Weak Scale Measurements? }}
\vskip 25pt
{\bf  Sacha Davidson \footnote{E-mail address:
davidson@thphys.ox.ac.uk}  and Alejandro Ibarra \footnote{E-mail address:
ibarra@thphys.ox.ac.uk} }
 
\vskip 10pt  
{\it Department of Physics, Theoretical Physics, University of Oxford \\
 1 Keble Road, Oxford, OX1 3NP, United Kingdom}\\
\vskip 20pt
{\bf Abstract}
\end{center}
\begin{quotation}
  {\noindent\small 
The see-saw mechanism is a very attractive explanation for small
neutrino masses, parametrized at the GUT scale
by the  right-handed
Majorana mass matrix, ${\cal M}$,
and the neutrino Yukawa matrix, ${\bf Y_\nu}$.
We show that in a SUSY model
with universal soft terms,   ${\cal M}$ and  ${\bf Y_\nu}$
can be calculated
from the light neutrino masses, the MNS matrix, 
and ${\bf Y^{\dagger}_\nu}
{\bf Y_\nu}$, which enters into the left-handed slepton radiative
corrections.
 This suggests that {\it in principle} the GUT-scale
inputs of the seesaw could be reconstructed from
the neutrino and sneutrino mass matrices. We
briefly discuss why this is impractical,
but advocate the neutrino and sneutrino mass
matrices as an alternative bottom-up parametrization
of the seesaw. 

\vskip 10pt
\noindent

}

\end{quotation}

\vskip 20pt  

\setcounter{footnote}{0}
\renewcommand{\thefootnote}{\arabic{footnote}}
\section{Introduction and notation}

The observed atmospheric \cite{atm1,atm2,K2K} and solar \cite{sol} neutrino
deficits suggest that
neutrinos have small but non-zero masses.
These can be elegantly
explained via the see-saw \cite{seesaw} mechanism,
where the left-handed neutrinos of the Standard Model
get a small mass from their small mixing
with the heavy right-handed singlet
neutrinos.  
Unfortunately, the new physics appears at a very high energy scale,
 not directly accessible to experiment, which
 makes it difficult to constrain the see-saw 
parameters.
Motivated by this, we will construct an alternative
parameter space, in terms of quantities measurable
(in principle) at low energies, where the experimental constraints
can be readily imposed. 

We consider the supersymmetric see-saw for two reasons:
first, supersymmetry stabilizes the Higgs mass against the dangerous quadratic
divergences that appear due to the presence of heavy particles (and in
the context of the see-saw mechanism, we know that there are  at least three
heavy particles, namely, the right-handed neutrinos). Second, the presence of
sleptons in the spectrum of the theory
is crucial to our derivation, because one of our low-energy
inputs is the slepton mass matrix.


The leptonic part of the superpotential reads
\bea
\label{superp}
W_{lep}= {e_R^c}^T {\bf Y_e} L\cdot H_1 
+ {\nu_R^c}^T {\bf Y_\nu} L\cdot H_2 
- \frac{1}{2}{\nu_R^c}^T{\cal M}\nu_R^c , \eea
where $L_i$ and $e_{Ri}$ ($i=e, \mu, \tau$) are the left-handed 
lepton doublet and the right-handed charged-lepton singlet, 
respectively, and $H_1$ ($H_2$) is the hypercharge $-1/2$ ($+1/2$)
 Higgs doublet.
 ${\bf Y_e}$ and ${\bf Y_{\nu}}$ are the Yukawa couplings that 
give masses to the charged leptons and generate the neutrino Dirac mass, 
and $\cal M$ is a $3 \times 3$ Majorana mass matrix that does 
not break the SM gauge symmetry. We do not  make any 
assumptions about the structure of the matrices in eq.(\ref{superp}), 
but consider the most general case. Then, it can be proved 
that the number of independent physical parameters is 21: 
15 real parameters and 6 complex phases \cite{santamaria}.

It is natural to assume that the  overall scale of $\cal M$, 
denoted by $M$, is much larger than the electroweak scale or any soft mass. 
Therefore, at low energies the right-handed neutrinos are decoupled and 
the corresponding effective Lagrangian reads
\bea  \delta {\cal L}_{lep}={e_R^c}^T {\bf Y_e} L\cdot H_1 
-\frac{1}{2}({\bf Y_\nu}L\cdot H_2)^T{\cal
M}^{-1}({\bf Y_\nu}L\cdot H_2) + {\rm h.c.},
\eea
After the electroweak symmetry breaking, the left-handed neutrinos
acquire a Majorana mass, given by
\bea
\label{seesaw}
{\cal M}_\nu= {\bf m_D}^T {\cal M}^{-1} {\bf m_D} =  {\bf Y_\nu}^T
{\cal M}^{-1} {\bf Y_\nu} \langle H_2^0\rangle^2,   \eea
suppressed with respect to the typical fermion masses by the inverse 
power of the large scale $M$. In what follows, it will be convenient 
to extract the Higgs VEV by defining 
\bea
\label{kappa}
\kappa = {\cal M}_\nu/ \langle H_2^0\rangle^2= {\bf Y_\nu}^T {\cal
M}^{-1} {\bf Y_\nu},  \eea
where $\langle H_2^0\rangle^2=v_2^2=v^2 \sin^2\beta$ and   $v=174$
GeV. Working in the flavour basis in which the charged-lepton Yukawa
matrix, $\bf{Y_e}$, and the gauge interactions are flavour-diagonal, the
$\kappa$ matrix can be diagonalized by the MNS \cite{MNS} matrix $U$ 
according to
\be
\label{Udiag}
U^T{\kappa } U={\mathrm diag}(\kappa_1,\kappa_2,\kappa_3)\equiv
D_\kappa, \ee
where $U$ is a unitary matrix that relates  flavour to mass eigenstates
\bea  \pmatrix{\nu_e \cr \nu_\mu\cr \nu_\tau\cr}= U \pmatrix{\nu_1\cr
\nu_2\cr \nu_3\cr}\,,
\label{CKM}
\eea
and the $\kappa_i$ can be chosen real and positive.  $U$ can be written as
\bea U=V\cdot {\mathrm diag}(e^{-i\phi/2},e^{-i\phi'/2},1)\ \ ,
\label{UV}
\eea
where $\phi$ and $\phi'$ are CP violating phases (if different from
$0$ or $\pi$) and $V$ has the ordinary form of the CKM matrix
\be \label{Vdef} V=\pmatrix{c_{13}c_{12} & c_{13}s_{12} & s_{13}e^{-i\delta}\cr
-c_{23}s_{12}-s_{23}s_{13}c_{12}e^{i\delta} & c_{23}c_{12}-s_{23}s_{13}s_{12}e^{i\delta} & s_{23}c_{13}\cr
s_{23}s_{12}-c_{23}s_{13}c_{12}e^{i\delta} & -s_{23}c_{12}-c_{23}s_{13}s_{12}e^{i\delta} &
c_{23}c_{13}\cr}.  \ee

The $\kappa$ matrix, eq.(\ref{kappa}), is at the moment our only 
experimental hint about the high energy physics that generates 
the neutrino masses. Unfortunately, it is not enough to 
reconstruct the whole theory. However, there is a second
window onto the high energy physics, apart from the neutrino
mass matrix: radiative corrections. 
Between the GUT scale and the Majorana mass scale, $M$, neutrino Yukawa 
couplings affect the renormalization of the slepton
mass matrix through the combination ${\bf Y^{\dagger}_{\nu}} {\bf Y_\nu}$. 
These contributions can leave  signatures at low energies and thus 
provide additional information about the theory at high energies. 
 With certain assumptions, 
the  information provided by radiative corrections 
is complementary to that provided by $\kappa$, so that {\it in principle}
it could be possible to reconstruct  the complete high energy theory. 
This is the case  if the soft masses are universal at
the high scale $M_X$, and if the most significant contributions to the
slepton RGEs come from the supersymmetric SM with
right-handed neutrinos.

There are many papers constraining
the seesaw parameter space, so we would like to
situate our work among this literature.
Present measurements of $\kappa$
do not determine the matrices ${\bf Y_{\nu}}$ and ${\cal M}$,
so seesaw analyses can be categorized by
what they use as inputs.
A common approach  is  to start at the
GUT scale, choose theoretically motivated textures
for ${\cal M}$ and the Yukawa matrices, require
that they induce a neutrino mass matrix consistent
with observations, and then study other low energy
predictions of the chosen texture. For recent
top-down discussions, see for instance \cite{LAU}
(about leptogenesis), \cite{topdown} (lepton flavour
violation) and references therein. 
A  more ``bottom-up''
approach \cite{bottomup},  is to start 
from the experimental data on neutrino masses
and mixings, and find some particular forms for ${\bf Y_{\nu}}$ 
and ${\cal M}$ that reproduce the neutrino data.
Some theoretical input is
required  at the GUT scale, because 
 there are fewer parameters in $\kappa$ than 
there are in  ${\bf Y_{\nu}}$ 
and ${\cal M}$.
 One can then  study other low energy predictions
of the chosen ${\bf Y_{\nu}}$ 
and ${\cal M}$.
Finally, in \cite{Casas:2001sr} the authors found a
parametrization of all the Yukawa couplings compatible
 with the low energy data, in terms of the masses of the 
right-handed neutrinos and some unknown parameters. This 
approach can be considered as ``hybrid'', since the
parameter space is described both by high energy and
low energy quantities.
The  bottom-up approach to the seesaw
which we follow, is to start from
present and hypothetical future
observations (of Supersymmetry), and try to
reconstruct  ${\cal M}$ and $ {\bf Y_\nu}$. 
All of our inputs will therefore be at the weak scale.

The paper is organized as follows. 
In section 2, we present the general procedure  to
reconstruct the high energy theory from the low energy inputs. 
It is evident that the more constrained the low energy parameters are, 
the better we know the high energy parameter space. Therefore, some
words on the present and future constraints on the low energy
observables are in order. This is done in section 3. Finally, 
in section 4, we discuss our results and some applications 
of our procedure.

\section{General procedure}

We take as our weak scale inputs the neutrino mass matrix, 
related to  $\kappa$,  and  
\beq
P \equiv {\bf Y^{\dagger}_{\nu}}  {\bf Y_\nu}~~.
\eeq 
At this stage, we do not have much information about $P$ 
and  we prefer to interpret it  as a way 
to parametrize our ignorance 
of the high energy physics. However, this parametrization will turn out 
to be very convenient, because $P$ 
 is the combination of Yukawa couplings which enters
in the slepton
Renormalization Group Equations (RGEs).

Now we turn to the determination of ${\bf Y_{\nu}}$ and ${\cal M}$ from 
$\kappa$ and $P$. We can always choose to work in the basis where 
the charged lepton mass matrix is diagonal. We can also choose 
to work in a basis of
right-handed neutrinos where ${\cal M}$ is diagonal
\be {\cal M}={\mathrm diag}({\cal M}_1,{\cal M}_2,{\cal M}_3)\equiv
D_{\cal M}, 
\ee
with ${\cal M}_i\geq 0$. In this basis, the neutrino Yukawa matrix must
be necessarily non-diagonal, but can always be diagonalized
by two unitary transformations:
\beq
\label{biunitary}
{\bf Y_\nu} = V_R^{\dagger} D_Y V_L.
\eeq
$V_L$ and $D_Y$ can be determined from $P$, since, using eq. (\ref{biunitary}),
\bea
\label{step1}
P \equiv  {\bf Y^{\dagger}_{\nu}}  {\bf Y_\nu}
 = V_L^{\dagger} D_Y^2 V_L . 
\eea
On the other hand,
 from $\kappa ={\bf Y}_\nu^{T}  D_{\cal M}^{-1} {\bf Y}_\nu $ and 
eq. (\ref{biunitary}),
\beq
\label{step2}
D_Y^{-1} V_L^* \kappa V^{\dagger}_L D_Y^{-1} = 
V_R^*  D_{\cal M}^{-1} V^{\dagger}_R,
\eeq
where the left hand side of this equation is known ($\kappa$ is one of 
our inputs, and $V_L$ and $D_Y$ were obtained 
from eq. (\ref{step1})). Therefore, $V_R$ and
 $D_{\cal M}$ can also be determined.
This shows  that, working in the basis where the charged 
lepton Yukawa coupling, ${\bf Y_e}$,
the right-handed Majorana mass matrix, ${\cal M}$, and the gauge
interactions are all diagonal, it is possible 
to determine {\it uniquely}
 the  heavy Majorana mass matrix, ${\cal M}$, and the neutrino Yukawa 
coupling, ${\bf Y}_\nu = V_R^{\dagger}  D_Y V_L$,
starting  from $\kappa$ and ${\bf Y}^{\dagger}_{\nu} {\bf Y_{\nu}}$. 
Notice  that the see-saw formula, eq. (\ref{kappa}),
is only valid at the Majorana mass scale and not at low energies, so
all the parameters in eqs.(\ref{step1}) and (\ref{step2}) should be 
understood at $M$.  Therefore, the observed 
$\kappa$ should be run from the electroweak
scale to the Majorana mass scale with the corresponding RGEs \footnote{This 
seems to require knowing the Majorana mass scale from the beginning, however,
in a numerical calculation, $M$ can be computed recursively.} \cite{rges}.

At this point it is worth checking that the number of physical
parameters is identical at high and low energies. If we want to reconstruct
the whole theory from low energy data, we must have 15 real parameters and 
6 complex phases at low energies, and this is actually the case. 
In the basis we have chosen to work in,
$\kappa$ contains six real parameters and three complex phases, 
${\bf Y_e}$ is determined by  three real parameters, 
and $P$ contains six real parameters and three 
complex phases, which add up to 15 real parameters and 6 complex phases. It 
is not immediately obvious that they are all independent,
but we have proved that indeed they are, calculating 
${\bf Y_{\nu}}$ and ${\cal M}$ explicitly from $\kappa$ and $P$. 

Our procedure shows that there is a one to one correspondence 
between the two following sets of parameters: 
$\{ {\bf Y_{\nu}},{\cal M} \}$ and $\{ \kappa,P \}$, and we are
free to pass from the former to the latter. A region in
any of them could be mapped onto the other without loss of 
information. The mapping 
$\{ {\bf Y_{\nu}},{\cal M} \} \rightarrow \{ \kappa,P \}$ corresponds to
the common top-down approach that has been extensively used in
the literature. Starting from theoretically motivated
textures for ${\bf Y_{\nu}}$ and ${\cal M}$, one can find the 
corresponding  $\kappa$ and $P$ and then check which of them are
consistent with the neutrino data and the bounds on lepton flavour
violation. 

The inverse mapping, 
$ \{ \kappa,P \} \rightarrow \{ {\bf Y_{\nu}},{\cal M} \} $,
corresponds to a bottom-up approach, which
we find appealing both for theoretical and
phenomenological reasons.  $\kappa$ and $P$
 contain the same information
about the see-saw as  ${\cal M}$ and  ${\bf Y}_\nu$,
but $\kappa$ is expressed in terms of
observable neutrino masses,
mixing angles and phases; and $P$ can in
principle be extracted  from renormalization
group effects.  It is therefore
straightforward to
restrict $\kappa$
and $P$ matrix elements to
lie within their experimentally allowed
ranges. The
experimentally allowed ${\cal M}$ and  ${\bf Y}_\nu$
can then be reconstructed. 
Ideally, we would like to shrink the allowed regions
for  $\kappa$ and $P$ to a point, and thus determine 
precisely ${\cal M}$ and  ${\bf Y}_\nu$, but as we will see in 
the next section, this seems far from being attainable. We would
like to stress here that the main goal of our paper is to
find an alternative (low energy) description of the see-saw
parameter space and not to determine ${\cal M}$ and  ${\bf Y}_\nu$
from experiments.

This novel parametrization could be applied to the study of the cosmological
baryon asymmetry  generated
from the out-of-equilibrium decay of
the $\nu_R$s \cite{LAU}: for
certain choices of  $ {\bf Y}_\nu$
and  ${\cal M}$, a sufficiently
large lepton asymmetry is produced
in the decay of the  $\nu_R$s, and
subsequently reprocessed into a baryon asymmetry
by non-perturbative
Standard Model B+L violating processes.
Starting from
$\kappa$ and $P$, we can
readily calculate the
implications of experimental measurements for
the leptogenesis scenario,
and study the dependence of the CP violating
asymmetry in $\nu_R$ decay \footnote{The generated
lepton asymmetry also depends on
cosmological parameters,
so additional assumptions are required to
calculate the baryon asymmetry.}
on weak scale masses and couplings \cite{DIP}.

To conclude this section, we would like to compare this
result to the quarks, where the 
$u_R$ and the $u_L$  share a Dirac mass---there is no
undetermined GUT-scale  mass. The up quark masses
squared are proportional to  the eigenvalues of 
$ {\bf Y}_u^{\dagger} {\bf Y}_u$,
which is also the combination that appears in the RGEs. 
This is in contrast to the neutrino sector, where
the light neutrino mass matrix 
is $ \kappa = {\bf Y}_\nu^{T} {\cal M}^{-1}{\bf Y}_\nu$,
and the left-handed slepton RGEs depend on 
 $ P =  {\bf Y}_\nu^{\dagger} {\bf Y}_\nu$. So,
the sleptons provide additional information,
complementary to the lepton mass matrices.  In the
neutrino sector,  $\kappa$, rather than $ {\bf Y}_\nu^{\dagger}{\bf Y}_\nu$,
is directly measurable at the weak scale.
$ {\bf Y}_\nu^{\dagger}{\bf Y}_\nu$ contributes to
the renormalization group running from the
high energy scale $M_X$ to ${\cal M}_i$;
indeed, it is the high scale input which
(in conjunction with $\kappa$)
allows us to determine ${\cal M}$.

\section{Present and future constraints on $\kappa$ 
and ${\bf Y^{\dagger}_{\nu}}  {\bf Y_\nu}$ }

All the 
matrix elements of $\kappa$ are in principle measurable, 
because they are determined from
the light neutrino masses and the angles and
phases of the MNS matrix. Three masses, three mixing angles
and three phases are required to fully reconstruct
$\kappa$. Atmospheric neutrino data determine 
 the neutrino
mass difference $|m_3^2 - m_2^2|$,
 and the mixing angle
$\theta_{23}$.  
Solar data allow various values for $m_2^2 - m_1^2$
and $\theta_{12}$, although present data seem to favour 
the large angle MSW solution  \cite{Gonzalez-Garcia:2001sq}. Also, 
reactor experiments \cite{Apollonio:1999ae} 
constrain $\theta_{13}$ to be small.
  Upcoming and proposed  
experiments hope to  determine the solar solution, and measure the
 angle $\theta_{13}$, the
sign of $m_3^2 - m_2^2$, and the $CP$
violating phase $\delta$ \cite{Cetal}.  The overall scale
of the neutrino masses could be determined
from microwave background and large scale
structure observations if $\sum_i m_i \simgt .1$ eV \cite{max}---
alternatively one could take the largest
mass to be the largest mass difference. The two remaining
``Majorana'' phases $\phi$ and $\phi'$ contribute
 to $\Delta L= 2$ neutrino interactions,
such as neutrinoless double beta decay \cite{0nbb}, which
could measure a combination of $\phi$ and $\phi'$.
An additional $\Delta L = 2$ process would
have to be measured to determine $\phi$ and
$\phi'$ separately.

 We now turn to constraining $P$, which
appears in the RGE of the left-handed slepton soft mass matrix, ${\bf m}_L^2$.
Since $P$ is related to radiative corrections, it is expected 
that the absolute value of any of its elements is $\simlt 16 \pi^2$. 
To further constrain $P$, it is necessary to make an 
ansatz about the high energy physics, and in what follows 
 we make two assumptions. We use the RGEs of the MSSM with right
handed neutrinos (see, for example, appendix in \cite{Casas:2001sr}) 
in running from
the high scale $M_X$ to $M$,
and then the RGEs of the MSSM between $M$ and $M_W$.
 This means that we neglect the
effects of any other new particles
between $M_W$ and $M_X$, and that we are assuming
$M_i \ll M_X$, so that the large log component of the 
$\nu_R$-induced loops is the most significant part.
Secondly, we assume that
the slepton soft mass matrices at high energies are proportional
to the identity. This is  naturally obtained in some well motivated 
scenarios, for instance, minimal supergravity, dilaton-dominated SUSY
breaking or gauge-mediated SUSY breaking. Also, model independent
analyses of flavour changing processes suggest approximate universality
of the soft terms. 
In making these assumptions, we are imposing
some hypotheses on the high energy theory and thus our approach 
is not strictly bottom-up. However, these minimal assumptions
are made in most top-down analyses, and  are a
small price to pay for parametrizing the see-saw 
with low energy data. With these requirements, the low energy left-handed
slepton mass matrices read, in the leading-log approximation, 
\bea
\label{softafterRG} 
\left(m^2_{\tilde{\ell}, \snu}\right)_{ij} & \simeq & 
 \left({\rm diagonal\,\, part}\right)_{\tilde{\ell}, \snu}
-\frac{1}{8\pi^2}(3m_0^2 + A_0^2)
({\bf Y^{\dagger}_\nu})_{ik} ({\bf Y_\nu})_{kj} \log\frac{M_X}{M_k}\ ,
\eea
where ``diagonal-part'' includes the tree level soft mass matrix, 
the radiative corrections from gauge and charged lepton 
Yukawa interactions, and the mass contributions from F- and D-terms
(that are different for charged sleptons and sneutrinos).
The off-diagonal elements in eq. (\ref{softafterRG}) induce rare lepton
flavour violating processes, such as  $\mu \rightarrow e \gamma$. The present 
upper bounds on their branching ratios can be used to constrain 
the absolute values of the off-diagonal 
elements of $P = {\bf Y^{\dagger}_{\nu}}{\bf Y_{\nu}}$, 
especially $|({\bf Y^{\dagger}_{\nu}}{\bf Y_{\nu}})_{12}|$, up to a log
dependence on the right-handed Majorana masses. 
For example, model independent analyses obtain $(m^2_L)_{12} 
\lsim 20 \,{\rm GeV}^2$ for a slepton mass of $100 \,{\rm GeV}$ \cite{GMS}, 
which translates into $|P_{12}| \lsim 6 \times 10^{-3}$. 
The observation of lepton flavour violating processes would set
a lower bound on $|P_{ij}|$, $i \neq j$, for a certain choice of
supersymmetric parameters.

Sparticle production at future colliders offers the possibility of
further constraining the left-handed slepton  mass matrices and 
therefore the magnitudes and phases of the elements of $P$. The 
  production of sneutrinos and the observation of their decays could 
determine  their masses and couplings to the charged leptons,
which would constrain the
elements of $m^2_{L}$.
The diagonal elements of 
${\bf Y}_\nu^{\dagger}{\bf Y}_\nu$ contribute to the running of the
diagonal elements of the left-handed slepton soft mass matrix, 
as do the gauge and charged lepton Yukawa couplings. 
So  elements  $P_{ii}$
that are of order the gauge couplings could
be determined from the
sneutrino masses.For a hierarchical 
 ${\bf Y}_\nu^{\dagger}{\bf Y}_\nu$, this appears difficult even
in the case of $[{\bf Y}_\nu^{\dagger}{\bf Y}_\nu]_{33}$, as has been
carefully discussed in \cite{Baer}. In some areas
of SUSY parameter space, slepton oscillations\cite{sleposc}
and lepton flavour violating slepton decays\cite{HP}  could
 determine 
$[m^2_{\tilde{\ell}}]_{13}$ and $[m^2_{\tilde{\ell}}]_{23}$
more accurately than rare $\tau$ decays.

The matrices ${\bf m}^2_{\tilde{\ell},\snu}$ are hermitian, so their
off-diagonal elements are complex. Naively, from sneutrino
oscillation experiments one could hope  to extract the
three phases of  ${\bf m}^2_{\snu}$, however, 
it is likely that only one combination  
is measurable. In these experiments, 
a flavour eigenstate
sneutrino could be produced with
a lepton of one
flavour (say $e^+$), oscillate
among the different $\snu$ mass
eigenstates, and then
decay into a lepton of another
flavour ($e.g.~ \mu^-$). 
As discussed in \cite{AFHC}, this process
only involves one phase, which could
be measured (for sneutrino mass differences of order
the decay rate) in the asymmetry
between the observed number of 
$e^+ \mu^-$ and  $e^- \mu^+$. 
The reason why there is only
one phase is that the lepton number conserving \footnote{we
neglect for the moment the sneutrino ``Majorana''
mass matrix, appearing in  $[{m}_{\snu \snu}^2]_{ij} \snu_i \snu_j + h.c.$}
sneutrino mass matrix 
$[{\bf m}^2_{\snu}]_{ij}$
can be diagonalized,
 in the basis defined after eq. (\ref{kappa}),
 by $W{\bf m}^2_{\snu} W^{\dagger} = D_{{\bf m}^2}$.
The unitary matrix $W$ has
six phases, five of which can be rotated away by choosing
 the relative phases of the charged leptons and
sneutrinos. This removal of
phases is similar to that in the CKM matrix.
It is possible because
 the phases of the left-handed leptons 
are not fixed by making their mass matrix diagonal and 
real, so their two relative
phases can be chosen to remove two phases,
either in $W$ {\it or} in the MNS matrix. In the case
of sneutrino oscillations,
two of the relative phases of the left-handed leptons 
used to redefine $W$ will reappear in the MNS matrix, but this does
not have any physical effect since this experiment does not involve 
neutrinos. Thus, to measure the remaining two phases we
need 
an experiment probing the sneutrino-neutrino-neutralino
vertex, that depends on  both $W$ and the MNS matrix.
Experiments that involve the MNS matrix
are difficult to perform,
so  measuring these two phases does
not seem feasible. 
An alternative approach might be  to look for CP violation
in sneutrino-anti-sneutrino oscillations \cite{GH}.
 Sneutrinos can oscillate into
anti-sneutrinos  due to small  $\Delta L = 2$ sneutrino
masses  $[{m}_{\snu \snu}^2]_{ij} \snu_i \snu_j + h.c.$, 
which are the soft SUSY breaking analogy of
the neutrino Majorana masses:
$[{m}_{\snu \snu}^2]_{ij} \sim v_2^2 m_0 \kappa_{ij}$.
However, this process is only
observable in a small area of the SUSY parameter space
\cite{GH}, so it is unlikely that the phases 
could be extracted from 
these  phenomena.

\section{Discussion}

In the best of all supersymmetric worlds,
where soft terms are universal at the
GUT scale,  and all masses, couplings and phases
have been measured at the weak scale,  it is possible to
calculate the GUT-scale  inputs of the SUSY
see-saw from measured quantities. However,
 in practice  it would
be very difficult to make all the weak
scale measurements  required to determine
the neutrino Yukawa matrix, ${\bf Y}_\nu$,  and the
right-handed neutrino Majorana mass matrix, ${\cal M}$.  The magnitude of
the off-diagonal elements of ${\bf Y}_\nu^{\dagger}{\bf Y}_\nu$
induce lepton flavour violation in the
left-handed slepton mass matrices, so could be measurable.
The diagonal elements of ${\bf Y}_\nu^{\dagger}{\bf Y}_\nu$
are more difficult: they modify the diagonal elements of
the left-handed slepton soft mass matrix, so they would be
hard to disentangle from gauge and charged lepton Yukawa contributions to
the RG running. Finally,
most of the CP violating phases in
 ${\bf Y}_\nu^{\dagger}{\bf Y}_\nu$
and the MNS matrix seem practically
unattainable.

A less ambitious, and at this stage  more practical, 
interpretation of our results, is that it 
is possible to describe the see-saw 
parameter space using just inputs at the electroweak scale, which
have a very straightforward physical interpretation (neutrino masses 
and mixings, rates for rare flavour changing processes, 
sneutrino masses...). Since we make our experiments at the
electroweak scale, it is much more natural to use
a parameter space spanned by quantities measurable at the electroweak 
scale rather than at the GUT scale. 
This approach  has the
 immediate consequence that the see-saw parameter space 
 is already constrained by neutrino data and 
rare lepton decays, and might be further constrained in the future with
the advent of neutrino factories and LHC.
Therefore, model independent conclusions  about the
 seesaw  mechanism can be drawn more readily in this
bottom-up formulation than in a top-down approach, where 
a priori the parameter space  is not constrained at all!

Our parametrization is very convenient to look for ways 
to test the see-saw mechanism.
We have shown that  it
is difficult to rule out the see-saw mechanism 
from neutrino data and radiative effects on 
left-handed slepton soft mass matrices:
for {\it any} combination of 
low energy neutrino parameters
(encoded in $\kappa$) and left-handed slepton mass matrices
(that can be parametrized by $P$, as in eq.(\ref{softafterRG})),
it is {\it always} possible to find a neutrino Yukawa coupling 
and a Majorana mass matrix that reproduce
those low energy observables \footnote{Notice that 
this  is not  obvious
 in the top-down approach: for example, it is not
immediately clear  that  there is a region of the 
 ${\bf Y}_\nu$ and ${\cal M}$ parameter space 
consistent with  any experimentally 
determined  neutrino parameters and
with the upper bounds on rare lepton decays.
}.  If  those ${\bf Y}_\nu$ and ${\cal M}$ are not compatible 
with perturbativity or the experimental lower bounds on 
Majorana masses, it would be a blow against the see-saw.
Also, the observation of
lepton flavour violation not encoded in
$\kappa$ and $P$  (for instance from
the right-handed slepton mass matrix), would
be an indication for physics other than the 
see-saw with universal soft terms. 
In this parametrization, neutrino masses and
mixing angles, and radiative effects
on slepton masses, are inputs, and in this sense cannot
be regarded as predictions of the see-saw mechanism. 
Therefore, to test
the see-saw,  we need another low energy
effect  which is a consequence of it,
such as  the baryon asymmetry of the Universe. The 
CP asymmetry in right-handed neutrino decay is
indeed a prediction of the see-saw and,  when combined
with assumptions  about the early evolution
of the Universe, could be
 compared with the cosmological  baryon
asymmetry.

In summary, we have shown that
in a SUSY model with universal soft terms, 
 it is possible to
determine the GUT scale inputs of the see-saw from
parameters that are in some sense measurable at the
weak scale. The right-handed neutrino Majorana mass matrix, ${\cal M}$,
and the neutrino Yukawa matrix, ${\bf Y}_\nu$, 
can be calculated from the light
neutrino masses, the MNS matrix,
and  ${\bf Y}_\nu^{\dagger}{\bf Y}_\nu$
(which enters into the renormalization
group equations of the left-handed slepton soft mass matrix).
This does not conflict with
decoupling theorems because
 ${\bf Y}_\nu^{\dagger}{\bf Y}_\nu$
(whose indices are left-handed)
is effectively a high-scale input: it
contributes to the
slepton mass matrix via renormalization group running
at scales above $M$.
This matrix can be extracted from $m_{\snu}^2$ if the soft
masses are universal at the high scale $M_X$, and if 
the main contribution to
the RGEs below $M_X$ is from the MSSM particles
and the right-handed neutrinos.  
We briefly discussed
the (not very encouraging) prospects
of measuring the
magnitudes and phases of all
$\kappa$ and  ${\bf Y}_\nu^{\dagger}{\bf Y}_\nu$
matrix elements, in a 
model with universal soft terms. A more
realistic approach is
to impose all the available experimental
constraints on $\kappa$ and 
 ${\bf Y}_\nu^{\dagger}{\bf Y}_\nu$,
vary the matrix elements over
their experimentally allowed ranges,
and calculate the resulting allowed
values of ${\bf Y}_\nu$ and ${\cal M}$.
We will pursue this bottom-up approach
to the see-saw (and its
application to leptogenesis) in subsequent work \cite{DIP}.

\subsection*{Acknowledgments}
We would like to thank Gian Giudice, Howie Haber and Martin Hirsch
for useful conversations. We are grateful to Graham Ross for comments
 and his continuous encouragement. Finally, we are especially indebted to Alberto Casas for enlightening discussions and a careful reading of the manuscript.

\end{document}